# CRYSTAL-CHEMISTRY AND SHORT-RANGE ORDER OF FLUORO-EDENITE AND FLUORO-PARGASITE: A COMBINED X-RAY DIFFRACTION AND FTIR SPECTROSCOPIC APPROACH

GIANCARLO DELLA VENTURA[1,2], FABIO BELLATRECCIA[1,2], FERNANDO CÁMARA[3] AND ROBERTA OBERTI [4]

1) Dipartimento di Scienze, Università di Roma Tre, Largo S. Leonardo Murialdo 1, I-00146 Roma, Italy
2) INFN, Laboratori Nazionali di Frascati, Roma
3) Dipartimento di Scienze della Terra, via Valperga Caluso 35, I-10125 Torino
4) CNR-Istituto di Geoscienze e Georisorse, UOS Pavia, via Ferrata 1, I-27100 Pavia, Italy

**ABSTRACT**

In this study we address the crystal-chemistry of a set of five samples of F-rich amphiboles from the Franklin marble (USA), using a combination of microchemical (EMPA), SREF, and FTIR spectroscopy methods. The EMPA data show that three samples fall into the compositional field of fluoro-edenite (Hawthorne *et al.,* 2012), whereas two samples are enriched in high-charged C cations, and - although very close to the $^{C}R^{3+}$ boundary - must be classified as fluoro-pargasite. Mg is by far the dominant C cation, Ca is the dominant B cation (with $^{B}Na$ in the range 0.00-0.05 apfu, atoms per formula unit), and Na is the dominant A cation, with $^{A}\square$ (vacancy) in the range 0.07-0.21 apfu; $^{W}F$ is in the range 1.18-1.46 apfu. SREF data show that: $^{T}Al$ is completely ordered at the *T*(1) site; the *M*(1) site is occupied only by divalent cations (Mg and $Fe^{2+}$); $^{C}Al$ is disordered between the *M*(2) and *M*(3) sites; $^{A}Na$ is ordered at the *A*(*m*) site, as expected in F-rich compositions. The FTIR spectra show a triplet of intense and sharp components at ~ 3690, 3675, and 3660 $cm^{-1}$, which are assigned to the amphibole, and the systematic presence of two very broad absorptions at 3560 and 3430 $cm^{-1}$. These latter are assigned, on the basis of polarized measurements and FPA (focal plane array) imaging, to chlorite-type inclusions within the amphibole matrix. Up to eight components can be fitted to the spectra; band assignment based on previous literature on similar compositions shows that $^{C}Al$ is disordered over the *M*(2) and *M*(3) sites, thus supporting the SREF conclusions based on the <*M*-O> bond distance analysis. The measured frequencies of all components are typical of O-H groups pointing toward Si-O(7)-Al tetrahedral linkages, thus allowing to characterize the SRO (short-range-order) of $^{T}Al$ in the double chain. Accordingly, the spectra show that in the fluoro-edenite/pargasite structure, the T cations, Si and Al, are ordered in such a way that Si-O(7)-Si linkages regularly alternate with Si-O(7)-Al linkages along the double chain.

## INTRODUCTION

Edenite, ideally $NaCa_2Mg_5Si_7AlO_{22}(OH)_2$, is an end-member of the calcium amphibole subgroup (Hawthorne *et al.,* 2012), and hence an hydroxyl-bearing double-chain silicate. It was first described by A. Breithaupt in 1874 based on a specimen from Franklin Furnace, New Jersey, USA (Clark, 1993). However, neither that sample nor the one studied later by Palache (1935) from the Franklin marble falls within the compositional range of edenite, as defined by the succeeding IMA Subcommittees on amphiboles, i.e. then $^{B}Ca \geq 1.50$ apfu (atoms per formula unit), $^{A}(Na,K) \geq 0.50$ apfu, $7.50 < {}^{T}Si < 6.50$ apfu and presently $^{W}(OH,F)$ and $^{A}Na$ dominant, $^{B}Ca/^{B}(Ca+Na) \geq 0.75$, $0.00 < {}^{C}R^{3+} \leq 0.50$, (e.g. Hawthorne *et al*., 2012). In fact, re-examination of several samples from the marble quarries in the Franklin area, kept and labelled as edenite in various Mineralogical Museums all around the world, showed these to be pargasite, or edenitic magniesio-hornblende or pargasitic-hornblende (our unpublished work; M. Lupulescu, personal communication).

Several attempts have been made to grow edenite experimentally. However, in OH-bearing systems the synthesis of end-member edenite has never been successful (Gilbert *et al.,* 1982). Na *et al.* (1986) investigated the phase relations in the system edenite + $H_2O$ and edenite + excess quartz + $H_2O$. They found that, when quartz is not in excess, phlogopite is the stable phase over a wide range of conditions including the amphibole *P*-*T* field, thus suggesting that edenite is probably stable only under very high $a_{SiO2}$. Indeed, excess quartz in the system stabilizes the amphibole; however, the compositions obtained are systematically solid-solutions in the ternary tremolite–richterite–edenite. Della Ventura and Robert (unpublished data) failed to synthesize edenite at $T$ = 700°C, and $P_{H2O}$ = 1 kbar. Della Ventura *et al.* (1999) studied the richterite–pargasite join, and observed that the amphibole is substituted by a Na-rich mica when approaching the edenite stoichiometry. In contrast, fluoro-edenite has been easily obtained by several authors in OH-free systems. Kohn and Komeforo (1955) synthesized fluoro-edenite and boron-rich fluoro-edenite starting from a mixture of oxides, at $P$ = 1 atm and using the melting technique, with slow cooling from 1350°C. The wet-chemical analysis of the bulk run-product gave: $Na_{0.99}(Ca_{1.84}Na_{0.16})(Mg_{4.79}Al_{0.18})(Si_{7.12}Al_{0.88})O_{22}F_{2.15}$. Graham and Navrotsky (1986) obtained high-yield amphibole run products, with less than 5% pyroxene impurities along the fluoro-tremolite–fluoro-edenite join. Raudsepp *et al.* (1991) also failed to synthesize edenite in the OH-system, but obtained high fluoro-edenite yields (> 90%) at $P$ = 1 atm; in their case, electron microprobe analyses (EMPA) showed significant deviation toward tschermakite. Boschmann *et al.* (1994) synthesized fluoro-edenite (with $^{B}Mg$ contents close to 0.20 apfu) using the same technique (slow cooling from 1240°C) and characterized their product by X-ray single-crystal diffraction and EMPA. The crystals were strongly zoned, from a core with composition close to end-member fluoro-edenite to a rim with composition intermediate between fluoro-tremolite and fluoro-pargasite. X-ray single-crystal refinement (SREF) showed that $^{T}Al$ is strongly ordered at *T*(1), and suggested that $^{C}Al$ is strongly ordered at *M*(2), this latter being a conclusion that should be revised in the light of this work. Gianfagna and Oberti (2001) reported on the occurrence of fluoro-edenite very close to the ideal composition, $NaCa_2Mg_5Si_7AlO_{22}F_2$, within the cavities of an altered benmoreitic lava at Biancavilla (Catania, Italy). Oberti *et al.* (1997) characterised a synthetic Mn-rich fluoro-edenite [$Na_{1.06}(Ca_{1.41}Na_{0.12}Mn_{0.47})(Mg_{4.44}Mn_{0.41}Al_{0.15})(Si_{6.91}Al_{1.09})O_{22}F_{2.00}$], and commented on $Mn^{2+}$ partitioning. Later, Oberti *et al.* (2006) reported on the occurrence of $^{A}(Na_{0.74}K_{0.02})^{B}(Ca_{1.27}Mn_{0.73})^{C}(Mg_{4.51}Mn^{2+}_{0.28}Fe^{2+}_{0.05}Fe^{3+}_{0.03}Al_{0.12}Ti_{0.01})^{T}(Si_{7.07}Al_{0.93})O_{22}(OH)_2$ - at that time the new end member parvo-mangano-edenite but now simply $^{B}Mn$-rich edenite after Hawthorne *et al*. (2012) - in the Grenville Marble of the Arnold mine, Fowler, St. Lawrence Co., New York (USA). They examined the possible bond-valence configurations proposed for edenite by Hawthorne (1997), and showed that the presence of $^{B}Mn$ substituting for $^{B}Ca$ helps to stabilize the charge arrangement of edenite.

In summary, the rarity of natural amphiboles with the edenite composition, their constant enrichment in Fe or Mn, and the failure in obtaining synthetic analogues of edenite, suggest that this amphibole is probably not stable (Raudsepp *et al.,* 1991) as confirmed by bond-valence considerations (Hawthorne, 1997).

We underwent a systematic work aimed at obtaining new structural and crystal-chemical data on edenite and fluoro-edenite; this study is focused on selected fluoro-edenite and fluoro-pargasite samples from the Franklin marble obtained from various mineral collections. We use a multi-methodological approach which includes EMPA, single-crystal X-ray diffraction and refinement and FTIR spectroscopy. In the last two decades, FTIR spectroscopy in the OH-stretching region has been used extensively in the study of hydrous minerals (e.g., Libowitzky and Beran, 2004 and references therein), and proved to be a fundamental tool to characterize short-range order/disorder features in amphiboles (Hawthorne and Della Ventura, 2007). In this paper, FTIR spectroscopy allowed us also to address two still poorly explored issues in amphibole crystal-chemistry, namely the short range order (SRO) of Mg/Al at the *M*(1-3) sites in the ribbon of octahedra and of Si/Al at the *T*(1,2) sites in the double chain of tetrahedra.

## STUDIED SAMPLES

Table 1 lists the labels and occurrence of the studied samples. All amphiboles, except FK-1, which has been provided as an isolated crystal about 3x3 cm in size, were extracted manually from the host marble, where the amphibole is associated with calcite, phlogopite and humite group minerals. Fluoro-edenite and fluoro-pargasite generally occurs as well-developed crystals with a prismatic habit along *c* and a colour ranging from grey-green to greenish-brown. The maximum crystal size is about 1 cm along *c*.

## CHEMICAL COMPOSITION

WDS analysis was done using a Cameca SX50 at Istituto di Geologia Ambientale e Geoingegneria (IGAG), Università di Roma La Sapienza. Analytical conditions were 15 kV accelerating voltage and 15 nA beam current, with a 5 μm beam diameter. Counting time was 20 s on both peak and background. Used standards were: wollastonite (Si*K*α, TAP; Ca *K*α, PET), periclase (Mg*K*α, TAP), corundum (Al*K*α TAP), orthoclase (K*K*α, PET), jadeite (Na*K*α, TAP), magnetite (Fe*K*α, LIF), rutile (Ti*K*α, LIF), Mn, Zn and Cr metal (Mn*K*α, Zn*K*α, LIF; Cr*K*α PET), syntetic fluoro-phlogopite (F*K*α, TAP), and silvite (Cl*K*α, PET). Analytical errors are 1% rel. Data reduction was done with the PAP method (Pouchou and Pichoir, 1985).

The crystal-chemical formulae were calculated based on 24 (O,OH,F,Cl) and 2 (OH,F,Cl) apfu, as suggested by the structure refinement which excluded the presence of significant oxo component ($^{W}O^{2-}$). The final data are given in Table 2. Three samples (G2552, GE408 and G405) fall into the fluoro-edenite compositional field, defined by $0.0 < {}^{C}R^{3+} < 0.5$ apfu (Hawthorne *et al.,* 2012), whereas samples J9698 and FK-1 are enriched in high-charged cations at C, and hence - although very close to the ${}^{C}R^{3+}$ boundary - must be classified as fluoro-pargasite. Magnesium is by far the dominant C cation, with minor Fe and Al, and very minor Ti. Ca is the dominant B cation, $^{B}$Na being negligible (max 0.05 apfu in sample C2552, Table 2). Na is the dominant A cation, K being always lower than 0.2 apfu and vacancy ranging from 0.07 to 0.21 apfu (Table 2).

## SINGLE-CRYSTAL STRUCTURE REFINEMENT (SREF)

The studied crystals were selected from mineral separates, mounted on a Philips PW-1100 four-circle diffractometer, and examined with graphite-monochromatized Mo*K*α X-radiation; crystal quality was assessed *via* profile analysis of Bragg diffraction peaks. Unit-cell dimensions were calculated from least-squares refinement of the *d* values obtained from 50 rows of the reciprocal lattice by measuring the gravity centroid of each reflection and its corresponding antireflection in the θ range –30° to 30°.

Intensity data were collected in the range $2° < \theta < 30°$. Only the monoclinic pairs *hkl* and $h\bar{k}l$ were collected with the Philips PW1100 point detector. Intensities (*I*) were corrected for absorption, Lorentz and polarisation effects, and then averaged and reduced to structure factors (*F*).

Reflections with $I > 3\ \sigma_I$ were considered as observed during unweighted full-matrix least-squares refinement on *F*. Scattering curves for fully ionized chemical species were used at those sites where chemical substitutions occur; neutral *vs.* ionised scattering curves were used at the *T* and anion sites (except O(3)). Selected crystal data and refinement information are given in Table 3, selected bond distances and site occupancies are given in Table 4, and refined values of the atomic coordinates and displacement factors are given in Table 5 (deposited).

Inspection of these data show that: (i) $^{T}Al$ is completely ordered at the *T*(1) site; (ii) only two samples - C2552 and GE408 - show minimal contents of C cations at the B sites; (iii) the A cations tend to order at the *A*(*m*) site, as expected in F-rich compositions; (iv) the *M*(1) site is occupied only by divalent cations (Mg and $Fe^{2+}$); (v) the <*M*(3)-O> distance are always shorter than the <*M*(1)-O> distances and in three samples - C2552, GE408 and G405 - even shorter than those of the *M*(2) site, where high-charged cations are ordered in (OH,F)-dominant amphiboles. However, the presence of fluorine is known to shorten the <*M*(1)-O> and <*M*(3)-O> distances equally in amphiboles, and thus these results are compatible with $^{C}Al$ disorder between the *M*(2) and *M*(3) sites, which had been so far found only in high *mg#* (Mg number = $Mg/Mg+Fe^{2+}$) pargasite crystallized at high *T*, *P* conditions (Oberti *et al.,* 1995a). This issue required confirmation with an independent technique, such as FTIR spectroscopy, which could also help to identify other peculiarities of local order in fluoro-edenites, an perhaps shed further light on the reasons for the rarity of edenitic compositions. It is useful to remind the reader here that a value of <*M*(3)-O> shorter than <*M*(1)-O> had been onserved also in nearly stoichiometric fluoro-edenite from Biancavilla (Gianfagna and Oberti, 2001) as well as in the synthetic sample by Boschmann *et al*. (1994). In those cases, however, the absence of OH-stretching bands and hence the impracticability of FTIR analysis did not allow the authors to address and clarify the issue of $^{C}Al$ ordering.

## POWDER INFRARED SPECTRA

FTIR spectra on powdered samples were collected on a Nicolet Magna 760 spectrophotometer equipped with a KBr beam-splitter and a DTGS detector. The spectra were acquired in the range 4000-400 $cm^{-1}$, the nominal resolution is 4 $cm^{-1}$ and the final spectra are the average of 32 scans. Samples were prepared as KBr pellets mixing 7 mg of the powdered samples with 150 mg of KBr. In order to minimize interfering moisture in the pellets, the mineral + KBr powders were dried at 110°C for 48 hrs and then pressed at about 9 ton/$cm^2$ for 5 minutes. All disks were kept in an oven (110°C) for 48 hrs and analysed at room temperature.

The collected spectra consist of a triplet of sharp and intense bands at ~ 3690, 3675, and 3660 $cm^{-1}$, respectively (Figure 1). A minor absorption at ~ 3640 $cm^{-1}$ is also visible in sample GE408. Although great care was exerted during the sample preparation, a very broad absorption in the $H_2O$ region, extending from 3650 to 3000 $cm^{-1}$ is clearly observed in all spectra (Figure 1). This absorption consists of at least two broad components at 3560 and 3430 $cm^{-1}$, respectively (Fig. 1). We tried to remove such a broad absorption with repeated heating of the pellets and immediate collection of the spectra; however, we observed only a minor decrease in intensity of this absorption. We had therefore to conclude that the 3560-3430 $cm^{-1}$ components are a real feature of the spectra and are not due to moisture adsorbed on the disk. This point will be discussed below.

## SINGLE-CRYSTAL INFRARED SPECTRA AND FPA IMAGING

Single-crystal spectra were collected with a beam size of 100 μm on randomly oriented double-polished slabs, using a Nicplan IR microscope equipped with a nitrogen-cooled MCT (mercury-cadmium-telluride) detector. 128 scans were averaged for every spectrum; the nominal resolution is 4 $cm^{-1}$. Several spectra showed the 3560-3430 $cm^{-1}$ broad bands already observed in powder spectra; however, the intensity of these components was extremely variable among the crystals. Representative single-crystal FTIR spectra are displayed in Figure 2, selected such as to show, for each sample, the less intense absorption at < 3600 $cm^{-1}$. Notably, the broad absorption at 3560-3430 $cm^{-1}$ appears not only in turbid region of the crystal, but occasionally in optically clear parts as well. This point is shown in Figure 3, where several spectra collected in different crystal areas are shown; note that the size of the analysed point is proportional to the beam size. Comparison with Figure 1 shows that the patterns are very similar to those collected using KBr pellets; in particular, the same triplet of sharp bands at ~ 3690, 3675 and 3660 $cm^{-1}$ is observed, but the overall intensity of the broad absorptions at 3560-3430 $cm^{-1}$ is greatly reduced. From these observations we conclude that the triplet of sharp bands is due to the amphibole phase, while the 3560-3430 $cm^{-1}$ bands must be assigned to inclusions.

To identify the nature of the inclusions, we first collected polarized spectra on some grains (using a gold wire grid on ZnSe substrate IR polarizer), and observed that all bands, including the amphibole triplet and the two additional broad absorptions have variable intensity as a function of the orientation of the electrical vector **E** with respect to the crystals. These measurements were done on randomly oriented samples, and thus do not allow any conclusion concerning the orientation of the O-H absorber in the mineral. However, they allow one to exclude the assignment of the broad absorptions at < 3600 $cm^{-1}$ to randomly oriented fluid or solid inclusions, and rather suggest the presence of some extra phase with a specific crystallographic orientation within the amphibole matrix. With the aid of literature data on hydrated substances (e.g. Farmer, 1974), we could identify this extra phase as a chlorite-type phyllosilicate. Selected single-crystal slices were studied using a Bruker Hyperion 3000 FTIR microscope equipped with a 64x64 focal-plane array (FPA) of detectors and a 15x Cassegrain objective at the INFN Laboratory of Frascati (Rome). The nominal resolution was 8 $cm^{-1}$ and 128 scans were averaged for the final spectra. Figure 4a shows the image obtained at point 6 in Figure 3 by integrating the intensity in the whole 3720-3200 $cm^{-1}$ frequency range. It shows the presence of a highly hydrated phase within the fluoro-edenite host, which corresponds to the intermixed layer silicate identified above. Optical observations show that this phase is typically oriented along the cleavage planes of the amphibole. The FTIR spectra shows that the phyllosilicate has an absorption in the integrated range which is by far higher than in the fluoro-edenite matrix; this suggests that, while the host amphibole is F-rich (see above), the associated layer-silicate is OH-rich, indicating a strong preference of the hydroxyl component for the layer-silicate structure with respect to the edenite structure. Kock-Muller *et al*. (2004) observed a similar feature in the spectra of omphacitic clinopyroxenes of mantle origin from beneath the Siberian platform, and could assign them, using FTIR and analytical TEM, to micrometric-to-nanometric chlorite-type and amesite inclusions. It is worth noting that TEM data collected on chips extracted from several zones of the pyroxene crystals showed that these guest phyllosilicates were present in both inclusion-rich and optically clear parts of the pyroxenes (Kock-Muller *et al.,* 2004). Similar intergrowths were observed by Cámara (1995) at the TEM scale on samples of amphibole in metabasites from the Nevado-Filabride complex. Analytical Electron Microscopy (AEM) showed that the amphibole has compositions related to edenite with significant glaucophane components, and hosts chlorite-biotite-quartz intergrowths at the nanoscale. Similar alteration textures were described in micaschists from the same complex by Soto (1991), consisting in chlorite+paragonite retrograde alteration developing in fractures in ferro-glaucophane.

## SPECTRUM FITTING AND BAND ASSIGNMENTS

The digitized spectra of Figure 2 were fitted by interactive optimization followed by least-squares refinement (Della Ventura *et al.*, 1996); the background was treated as linear and all bands were modelled as symmetric Gaussians (Strens, 1974). The spectra were fitted to the smallest number of peaks needed for an accurate description of the spectral profile. The distribution of absorption, $y$, as a function of energy (wavenumber, $x$) was described by the relation $y = A \exp [- 0.5 (x – \mathrm{P} / \mathrm{W})^2]$ where $A$ is the amplitude, $P$ is the peak centroid, and $W$ is the full-width at half-maximum height (FWHM). The spectral parameters (position and width) of some strongly overlapping peaks were refined where these peaks are most prominent, and then fixed during the refinement of the other samples. This is clearly the case of the tremolite band at around 3670 $cm^{-1}$ (band T in Fig. 6) which was refined in samples G405 and G408 and than added to all other samples on the base of the EMPA + XRD results which indicated a systematic presence of $^{A}\square$ in the structures of the examined specimens. At convergence, the peak positions were released and the FWHM only were constrained to be roughly constant in all spectra.

A typical example of a fitted spectrum is shown in Figure 5, while all the relevant data are given in Table 6. The spectra of the studied samples consist of two doublets at 3703-3691 and 3675-3659 $cm^{-1}$ having similar relative intensity and frequency separation (~ 15 $cm^{-1}$), a sharp band at ~ 3672 $cm^{-1}$ and three rather broad bands at 3641-3622 $cm^{-1}$ and 3727 $cm^{-1}$ which account for the tails of the absorption.

Assignment of these bands can be based on previous studies on synthetic (OH,F) amphiboles; to this purpose, we compared in Figure 6 the spectrum of fluoro-edenite C2552 (this work) with the spectra of a fluoro-pargasite from Robert *et al.* (2000) and a fluoro-richterite from Robert *et al.* (1999), both of which have F = 1.2 and OH = 0.8 apfu.

The spectrum of synthetic fluoro-richterite (Fig. 6, bottom) shows a main band (A) at 3730 $cm^{-1}$ which is assigned to an O-H group bonded to a $^{M(1)}\mathrm{Mg}^{M(1)}\mathrm{Mg}^{M(3)}\mathrm{Mg}$ trimer of octahedra (Robert *et al.*, 1989; Della Ventura, 1992; Robert *et al.*, 1999), directed parallel to *a** and pointing toward $^{A}$Na when the A cavity is built by a ring of tetrahedra all occupied by Si. This configuration can be expressed as MgMgMg-OH-$^{A}$Na:SiSi when using the notation of Della Ventura *et al*. (1999) and Hawthorne *et al.* (2000). While OH is replaced by F, a new band appears in the spectrum at 3710 cm$^{-1}$ (A*); its position does not depend on the OH/F ratio, and its relative intensity varies with the F content. The reason for the observed downward shift of the OH-band due to the $OH_{-1}F$ substitution has been discussed in detail by Robert *et al.* (1999). It is based on the observation that in fluorine-rich amphiboles the A cation is displaced from the center of the cavity, along the mirror plane and toward the fluorine site (i.e., at the *A*(*m*) site), in agreement with the presence of *A*-F attraction (Hawthorne *et al.*, 1996a). This shift reduces the repulsive *A*-H interaction between the A cation and the hydrogen atom of the opposite O-H group (Fig.7). Thus, the O-H group in the local OH-*A*-F configuration absorbs the IR radiation at a lower frequency than in the OH-*A*-OH configuration.

Finally, the minor absorption at 3672 cm$^{-1}$ indicates a slight but significant departure of the composition toward tremolite (Robert *et al.,* 1989; Hawthorne *et al*., 1997; Gottschalk *et al.,* 1999), and is assigned to configurations including an empty A site (~ 5%) observed in the sample.

The spectrum of end-member pargasite has been previously studied by several authors (Semet, 1973; Della Ventura *et al.,* 1999, Della Ventura *et al.,* 2003); it shows a doublet of rather broad bands with almost equal intensity, centered at 3709 and 3678 cm$^{-1}$, respectively. Using the same band nomenclature, these are assigned to the local configurations MgMgMg-OH-$^{A}$Na:SiAl and MgMgAl-OH-$^{A}$Na:SiAl, respectively, and derives from nearly complete Al disorder between the *M*(2) and *M*(3) sites (Oberti *et al*., 1995a). Note that the difference in wavenumber (~ 30 cm$^{-1}$) between these two bands is due solely to the different nearest-neighbour (NN) octahedral environment ($^{M(1)}$Mg$^{M(1)}$Mg$^{M(3)}$Mg *vs* $^{M(1)}$Mg$^{M(1)}$Mg$^{M(3)}$Al) of the OH group, the rest of the structure in the second (*T* sites) and third (*A* site) coordination shell being unchanged. Note also that in pargasite, the band assigned to the vibration of the O-H group bonded to a $^{M(1)}$Mg$^{M(1)}$Mg$^{M(3)}$Mg trimer of octahedra is shifted by 20 cm$^{-1}$ with respect the same band in richterite; the reason has been explained by Della Ventura *et al.* (1999), and is related to the presence of Si-O(7)-Al linkages in the rings of tetrahedra in pargasite. This point will be discussed in more detail later. Similar to richterite, the presence of F at the O(3) site in pargasite shifts both the bands at 3709 and 3678 cm$^{-1}$ (A and B, respectively, in Fig. 6, top) by 15 cm$^{-1}$ downwards (A* and B*), and the intensity of the new bands depends on the F content (Robert *et al.,* 2000).

The spectrum of fluoro-edenite (Fig. 6, middle) is similar to that of fluoro-pargasite, and its components can be assigned in a similar way. In particular, the A and B component can be assigned to the MgMgMg-OH-$^{A}$Na:SiAl and MgMgAl-OH-$^{A}$Na:SiAl configurations, while the A* and B* components can be assigned to the MgMgMg-OH/F-$^{A}$Na:SiAl and MgMgAl-OH/F-$^{A}$Na:SiAl configurations. The T component can be assigned to the tremolite-type local environment, i.e. to presence of $^{A}\square$, in accordance with the microchemical data of Table 2. Finally, the lower frequency bands G and H, at 3641and 3622 cm$^{-1}$, respectively, are assigned to configurations involving Al occurring both as C and T cation and coupled with $^{A}\square$, i.e. to magnesio-hornblende-type configurations (Hawthorne et al., 2000). Note that a weak component around 3720 cm$^{-1}$ is required in two samples to model the peak asymmetry on the high-frequency side of the main band. The intensity of this component is however so low to be neglected in the above discussion. Final band assignments are summarized in Table 7

In conclusions, FTIR spectra clearly show that (i) $^{C}$Al in the studied samples is disordered over the M(2) and M(3) sites, thus confirming the SREF conclusions based on the analysis of the <*M*-O> bond distances; (ii) similar to pargasite, the measured frequencies of all OH bands (except the minor "tremolite”-type component at ~3672 cm$^{-1}$) are typical of O-H groups pointing toward Si-O(7)-Al tetrahedral linkages.

A point which is worth discussing here is the possible use of FTIR spectra for quantitative purposes. There are several problems involved in this issue, and some of these have been discussed by Hawthorne *et al.* (1996b, 2000). The key factors in this regard are: (i) the obvious difficulty in using intensity data (usually band areas) obtained after a decomposition process, which might be highly subjective in the case of severe band overlap, and (ii) the relationship between the absorption factor (ε) and the wavenumber (ν) (e.g. Libowitzky and Rossman, 1997 for hydrous minerals in general, and Skogby and Rossman, 1991 for amphiboles, in particular) which is still not completely understood (Della Ventura *et al.*, 1996, Hawthorne *et al.*, 1996b). Due to these problems, the measured intensities cannot be converted accurately to site-populations. For the present case, the site-populations derived from the spectral decomposition and band assignment must be in accord with the data derived by SREF on the same samples. If this is the case, the band parameters such as the band width can be confidently constrained to appropriate values and then compared with the data known from the refinement of synthetic samples (e.g. Della Ventura *et al.*, 1999). We can test the band fitting model described above by calculating the OH/F composition and the amount of A cations from the intensity data of Table 6. For the OH/F ratio we used the relationship $I_A+I_B/(I_A+I_{A*}+I_B+I_{B*})$ (Robert *et al.*, 2000), while for the A cations we used the equation $x = R/[k+R(1-k)]$ (Hawthorne *et al.*, 1997), where $x$ is the tremolite component in the amphibole, $R$ is the relative intensity ratio between $A$-site vacant and $A$-site occupied OH-bands and $k$ is 2.2. The results given in Table 8 show a good agreement between FTIR- and EMPA-derived site populations, and suggest that both the band assignment and the fitting model are adequate, at least for well-characterized and sufficiently chemically simple amphiboles.

## SHORT-RANGE ORDER OF $^{T}$AL IN FLUORO-EDENITE AND FLUORO-PARGASITE

The knowledge of Al partitioning in amphiboles has important implications for the geobarometry of igneous and metamorphic processes (e.g., Spear, 1981; Hammarstrom and Zen, 1986; Hollister *et al.,* 1987, Ridolfi *et al.*, 2010). Most thermodynamic analyses (e.g., Graham and Navrotsky, 1986) have so far considered activity models based solely on the occurrence of $^{C}$Al at $M$(2) and $^{T}$Al at $T$(1). Crystal-chemical work has shown that these mixing models are inadequate, at least in pargasite, where $^{C}$Al can disorder between the $M$(2) and $M$(3) sites in Mg-rich compositions crystallised at high $T$ (e.g., Oberti *et al.*, 1995a) and $^{T}$Al can either occupy the $T$(2) site at $^{T}$Al > 2.0 apfu or disorder between the $T$(1) and $T$(2) sites at high temperature (Oberti *et al.*, 1995b). Long-range ordering (LRO) patterns of Al in amphiboles are now understood quite well (see Oberti *et al.,* 2007 for a complete discussion of modern data), but we still know very little about short-range order of Al. Because the X-ray scattering factors for Al and Si are very similar, X-ray diffraction can provide long-range information on the ordering pattern of T cations based solely on the analysis of the mean <$T$(1)-O> and <$T$(2)-O> bond distances. When applied to amphiboles, the bond-valence theory (Brown, 1981, 2002) systematically showed that cation ordering patterns are strongly constrained by the bond-strength requirements of the coordinated oxygen atoms (Hawthorne, 1997). One resulting and fundamental feature, in amphibole crystal-chemistry is the avoidance of Al-O-Al linkages in the double chain of tetrahedra. Attempts of understanding both long-range and short-range order via $^{29}$Si MAS NMR, which is sensitive to NN (nearest neighbour) and NNN (next nearest neighbour) environments around the target Si nuclei, have been done by Welch *et al.,* (1994, 1998). However, the $^{29}$Si MAS NMR results are still relatively difficult to interpret and model.

A deeper insight into this problem can be obtained by considering the stereochemistry of the *T* sites in Al-bearing amphiboles, and by close comparison of the OH-spectra of fluoro-edenite with end-members richterite and pargasite (Fig. 8).

In the fluoro-edenite and fluoro-pargasite samples of this work, structure refinement results indicate that $^{T}$Al is completely ordered at *T*(1). As stated above, the higher-frequency component in the spectrum of pargasite is assigned to the same NN configuration of richterite, i.e. *M*(1)*M*(1)*M*(3)-O(3)-*A* = MgMgMg-OH-$^{A}$Na. The observed shift in pargasite toward lower frequency with respect to richterite (3709 cm$^{-1}$ *vs* 3730 cm$^{-1}$, Raudsepp *et al.,* 1987, Della Ventura *et al.,* 1999, 2001) is an important feature in the FTIR spectroscopy of the amphiboles (and any hydroxyl-bearing silicate). This has been explained by an hydrogen bond to the O(7) oxygen atom (Fig. 9), the strength of the bond controlling the shift in the frequency of the principal OH-stretching vibration (Della Ventura *et al.,* 1999, 2003; Libowitzky and Beran, 2004). The key issue here is that the valence-sum rule must be satisfied at the O(7) anion, which is bonded solely to two *T*(1) tetrahedra (Fig. 9). When the *T*(1)-O(7)-*T*(1) linkages are of the type Si-O(7)-Al, the deficit in bond valence at the O(7) anion must be alleviated *via* a stronger O(3)-H…O(7) bonding. Hence we can use the spectra of Figure 1 and 2 to evaluate the SRO of cations at the *T*(1) sites in the studied amphiboles.

Consider the double-chain of tetrahedra in richterite (Fig. 10a). All the tetrahedra are occupied by Si, and hence all *T*(1)-O(7)-*T*(1) linkages are of the Si-O(7)-Si type. Therefore, we observe a unique OH-stretching band at 3730 cm$^{-1}$. Consider next the double-chain of tetrahedra in stoichiometric pargasite (Fig. 10b). The composition of the double chain of tetrahedra is $Si_6Al_2$; hence, the *T*(2) sites are occupied by Si and half of the *T*(1) sites are occupied by Al. All the *T*(1)-O(7)- *T*(1) linkages must be of the Si-O(7)-Al type to avoid Al-O-Al linkages (cf. the Löwenstein rule; Löwenstein, 1954). In such a case, all the H atoms are involved in an hydrogen bonding with the closest O(7) atom, and all the OH-stretching bands are displaced by 20 cm$^{-1}$ toward lower frequency. In particular, the MgMgMg-OH-$^{A}$Na:SiAl band is found at 3709 cm$^{-1}$. In local configurations where $^{C}$Al occurs at the NN *M* sites, i.e. the MgMgAl-OH-$^{A}$Na:SiAl configurations, the band has an additional shift of ~30 cm$^{-1}$, i.e. at 3675 cm$^{-1}$ with respect to the corresponding band in richterite (band B in Fig. 6).

Consider now fluoro-edenite. In stoichiometric edenite, the composition of the double chain of tetrahedra is $Si_7Al_1$ hence half of *T*(1)-O(7)- *T*(1) linkages must be of the Si-O(7)-Si type, and half must be of the Si-O(7)-Al type; in such a case, there are at least two possible patterns of order between $^{T(1)}$Si and $^{T(1)}$Al, and these must have different spectral expressions in the infrared region. Figure 10c shows the most ordered pattern, with Si-O(7)-Si linkages regularly alternating with Si-O(7)-Al linkages. Figure 10d shows another "ordered" possibility, where clusters of Si-O(7)-Al linkages alternate with clusters of Si-O(7)-Si linkages (the number of the linkages in the two clusters being equal).

In the first model, all the OH-stretching bands must be of the pargasite-type and fall at ~ 3709 $cm^{-1}$ [or at lower wavenumber if cations different from Mg occur at the *M*(1,3) sites, e.g. Della Ventura *et al*., 1996]. In the second model, we should observe two bands with almost the same intensity occurring at 3730 and 3709 $cm^{-1}$ [again, if the C cation composition is $Mg_5$]. The systematic lack of a band at 3730 $cm^{-1}$ in the spectrum of the studied fluoro-edenite and fluoro-pargasite samples of this work (Figure 1,2) shows that this is not the case. Hence, the double chain of tetrahedra in fluoro-edenite has the SR ordered configuration schematically shown in Figure 10c.

## ACKNOWLEDGMENTS

Marcello Serracino assisted during the EMP analyses and Antonio Gianfagna supported financially the EMP analyses. Thanks are due to Dr. Pete J. Dunn, Smithsonian Institution, National Museum of Natural History, and Dr. John Cianciulli, Franklin Mineral Museum for providing the studied samples. RO acknowledges funding from the MIUR-PRIN 2009 project "Structure, microstructures and cation ordering: a window on to geological processes and geomaterial properties". Thanks are due to S. Mills, D. Jenkins and a further anonymous referee for their helpful suggestions.

## Figure captions

**Fig. 1** IR OH-stretching powder-spectra for the studied fluoro-amphiboles.

**Fig. 2** Single-crystal IR OH-stretching spectra for the studied fluoro-amphiboles. Absorbance scale normalized to thickness.

**Fig. 3** Variation of the OH-stretching spectrum as a function of the analysed point. Sample C2552, thickness 174 μm.

**Fig. 4** Selected FPA image (b) of sample C2552; the corresponding optical image is given in (a). The scale color from blue (minimum) to red (maximum) is proportional to the intensity in the water stretching 3720 - 3200 $cm^{-1}$ region.

**Fig. 5** The fitted spectrum of sample C2552; open squares: experimental pattern, broken lines: fitted bands, line: resulting envelope.

**Fig. 6** The fitted spectrum of Figure 5 (middle) compared with the fitted spectra of synthetic fluoro-pargasite (top, from Robert *et al.*, 2000) and synthetic fluoro-richterite (bottom, from Robert *et al.*, 1999), both the synthetic samples samples having F = 1.2 apfu.

**Fig. 7** A sketch of the *C*2/*m* amphibole structure showing the local environment around the O-H dipole.

**Fig. 8** The OH-stretching spectra of synthetic pargasite (from Della Ventura *et al.*, 1999), fluoro-edenite from Franklin (sample C2552, this work), and synthetic richterite (from Robert *et al.*, 1989). The local configurations are schematically given in the figure.

**Fig. 9** The local environment of the H atom in the *C*2/*m* amphibole structure view down [010]; the *T*(2) sites have been omitted for clarity. Green = Si, grey = Al. Modified from Della Ventura *et al.* (1999).

**Fig. 10** Schematic tetrahedral chain arrangements in (a) richterite; (b) pargasite; (c) edenite, regularly alternated; (d) edenite, clustered. For more explanation, see text. Blue tetrahedra = Si, orange tetrahedra = Al. Note that when the *T*(1)-O(7)-*T*(1) linkages have the configuration Al-O(7)-Si there can be orientational disorder relative to the *b* direction.

**Table 1** Sample labels and occurrence of the samples of this work

| Sample | Occurrence | Name | IGG code |
|---|---|---|---|
| J9698 | Edenville, Orange Co., New York, USA; SI | Fluoro-pargasite | 1073 |
| C2552 | Sterling Hill, Ogdensburg, New Jersey, USA; SI | Fluoro-edenite | 1069 |
| GE408 | Limecrest-Southdown Quarry, Sparta, New Jersey, USA; FM | Fluoro-edenite | 1070 |
| G405 | Franklin Quarry, Franklin, New Jersey, USA; FM | Fluoro-edenite | 1071 |
| FK1 | Edenville, Orange Co., New York, USA; FM | Fluoro-pargasite | 1072 |

SI: Smithsonian Institution; FM: Franklin Mineral Museum

**Table 2** Microchemical data and crystal-chemical formulae for the samples of this work.

| | J9698 4 an. | C2552 5 an. | GE408 3 an. | G405 6 an. | FK1 3 an. |
|---|---|---|---|---|---|
| $SiO_2$ | 47.42 | 47.83 | 50.23 | 48.37 | 46.59 |
| $TiO_2$ | 0.52 | 0.38 | 0.07 | 0.09 | 0.38 |
| $Al_2O_3$ | 11.34 | 7.29 | 5.81 | 9.40 | 11.14 |
| $Cr_2O_3$ | 0.03 | 0.03 | 0.03 | 0.03 | 0.02 |
| FeO | 1.83 | 4.10 | 1.51 | 1.69 | 3.10 |
| $Fe_2O_3$ | 0.00 | 0.45 | 0.00 | 0.00 | 0.05 |
| MnO | 0.03 | 0.09 | 0.08 | 0.02 | 0.02 |
| MgO | 20.26 | 20.30 | 22.57 | 20.90 | 19.37 |
| ZnO | 0.01 | 0.05 | 0.03 | 0.05 | 0.11 |
| CaO | 13.48 | 12.51 | 13.36 | 13.09 | 13.15 |
| $Na_2O$ | 2.95 | 3.33 | 2.37 | 2.69 | 2.76 |
| $K_2O$ | 0.64 | 0.28 | 0.61 | 1.11 | 0.72 |
| F | 3.20 | 2.89 | 2.55 | 3.29 | 2.63 |
| Cl | 0.04 | 0.05 | 0.03 | 0.03 | 0.11 |
| $H_2O^*$ | 0.63 | 0.71 | 0.91 | 0.57 | 0.85 |
| | 102.38 | 100.29 | 100.16 | 101.34 | 100.98 |
| O=F,Cl | -1.36 | -1.23 | -1.08 | -1.39 | -1.13 |
| Total | 101.02 | 99.06 | 99.08 | 99.94 | 99.85 |
| | | | | | |
| Si | 6.60 | 6.86 | 7.09 | 6.80 | 6.60 |
| Al | 1.40 | 1.14 | 0.91 | 1.20 | 1.40 |
| ∑T | 8.00 | 8.00 | 8.00 | 8.00 | 8.00 |
| | | | | | |
| Al | 0.46 | 0.09 | 0.06 | 0.36 | 0.46 |
| $Fe^{3+}$ | 0.00 | 0.05 | 0.00 | 0.00 | 0.00 |
| Cr | 0.00 | 0.00 | 0.00 | 0.00 | 0.00 |
| Ti | 0.05 | 0.04 | 0.01 | 0.01 | 0.04 |
| Zn | 0.00 | 0.01 | 0.00 | 0.01 | 0.01 |
| Mg | 4.21 | 4.34 | 4.75 | 4.38 | 4.09 |
| $Fe^{2+}$ | 0.21 | 0.49 | 0.18 | 0.20 | 0.37 |
| $Mn^{2+}$ | 0.00 | 0.01 | 0.01 | 0.00 | 0.00 |
| ∑C | 4.94 | 5.03 | 5.01 | 4.97 | 4.98 |
| | | | | | |
| ΔC | - | 0.03 | 0.01 | - | - |
| Ca | 2.00 | 1.92 | 1.99 | 1.97 | 2.00 |
| Na | 0.00 | 0.05 | 0.00 | 0.03 | 0.00 |
| ∑B | 2.00 | 2.00 | 2.00 | 2.00 | 2.00 |
| | | | | | |
| Ca | 0.01 | 0.00 | 0.03 | 0.00 | 0.00 |
| Na | 0.80 | 0.88 | 0.65 | 0.71 | 0.75 |
| K | 0.11 | 0.05 | 0.11 | 0.20 | 0.13 |
| ∑A | 0.92 | 0.93 | 0.79 | 0.91 | 0.88 |
| | | | | | |
| OH | 0.58 | 0.68 | 0.86 | 0.53 | 0.80 |
| F | 1.41 | 1.31 | 1.14 | 1.46 | 1.18 |
| Cl | 0.01 | 0.01 | 0.01 | 0.01 | 0.03 |
| ∑W | 2.00 | 2.00 | 2.00 | 2.00 | 2.00 |

∗ calculated; formula normalised to 22 oxygen atoms and 2 (OH,F,Cl).

**Table 3** Unit-cell dimensions and crystal-structure refinement information

| Sample | J9698 | C2552 | GE408 | G405 | FK1-2 |
|---|---|---|---|---|---|
| $a$ (Å) | 9.849(2) | 9.865(4) | 9.871(3) | 9.866(4) | 9.862(2) |
| $b$ (Å) | 17.973(4) | 18.026(7) | 18.033(9) | 17.987(11) | 17.998(5) |
| $c$ (Å) | 5.2867(14) | 5.2836(16) | 5.2805(18) | 5.2817(25) | 5.2942(17) |
| $\beta$ (°) | 105.17(2) | 105.013(24) | 104.97(3) | 105.16(4) | 105.17(2) |
| $V$ (Å$^3$) | 903.26 | 907.50 | 908.02 | 904.65 | 906.95 |
| | | | | | |
| $F_{all}$ | 1498 | 1501 | 1499 | 1492 | 1375 |
| $F_{obs}$ $I > 3\sigma_I$ | 1346 | 1342 | 1326 | 1325 | 1212 |
| | | | | | |
| $R_{sym}$ | 1.80 | 1.10 | 1.20 | 1.60 | 4.30 |
| $R_{obs}$ $I > 3\sigma_I$ | 1.84 | 1.42 | 1.29 | 1.46 | 1.77 |
| $R_{all}$ | 2.34 | 1.88 | 1.68 | 1.90 | 1.69 |
| | | | | | |
| Size (mm) | 0.73 x 0.50 x 0.50 | 0.63 x 0.36 x 0.36 | 0.51 x 0.30 x 0.26 | 0.66 x 0.53 x 0.50 | 0.38 x 0.33 x 0.23 |
| θ-range | 2-30° | 2-30° | 2-30° | 2-30° | 2-30° |
| kV | 40 | 50 | 50 | 40 | 50 |
| mA | 20 | 25 | 20 | 20 | 30 |

**Table 4** Selected geometrical parameters (Å) and refined site-scattering values (ss, epfu)

| Sample | J9698 | C2552 | GE408 | G405 | FK1-2 |
|---|---|---|---|---|---|
| <*M*(1)-O> | 2.068 | 2.070 | 2.069 | 2.067 | 2.075 |
| <*M*(2)-O> | 2.051 | 2.077 | 2.079 | 2.061 | 2.054 |
| <*M*(3)-O> | 2.055 | 2.055 | 2.056 | 2.054 | 2.061 |
| <*M*(4)-O> | 2.490 | 2.502 | 2.505 | 2.496 | 2.493 |
| <*A*-O> | 2.927 | 2.923 | 2.931 | 2.932 | 2.934 |
| <*A*(*m*)-O> | 2.873 | 2.862 | 2.870 | 2.879 | 2.882 |
| <*A*(2)-O> | 2.630 | 2.614 | 2.630 | 2.641 | 2.632 |
| <*T*(1)-O> | 1.660 | 1.653 | 1.646 | 1.654 | 1.662 |
| <*T*(2)-O> | 1.634 | 1.632 | 1.633 | 1.633 | 1.634 |
| | | | | | |
| ss *M*(1) | 25.01 | 25.82 | 24.52 | 24.75 | 25.75 |
| ss *M*(2) | 26.79 | 28.56 | 25.22 | 25.97 | 27.38 |
| ss *M*(3) | 12.63 | 12.93 | 12.26 | 12.46 | 13.03 |
| ss *M*(4) | 40.09 | 39.04 | 39.53 | 39.92 | 40.06 |
| ss *M*(4') | - | 0.90 | 0.59 | - | - |
| ss *A* | 2.37 | 2.04 | 1.74 | 2.68 | 2.47 |
| ss *A*(*m*) | 5.43 | 5.12 | 4.68 | 5.51 | 5.26 |
| ss *A*(2) | 3.12 | 2.76 | 2.40 | 2.71 | 2.35 |

**Table 6** Positions ($cm^{-1}$), widths ($cm^{-1}$), absolute and relative intensities for the bands A-H in the infrared OH-stretching spectra. The thickness of each studied section is also given. Note that a weak component around 3720 $cm^{-1}$ is required in two samples to model the peak asymmetry on the high-frequency side of the main band. The intensity of this component is however so low to be neglected.

| Band | Parameter | J9698 | C2552 | GE408 | G405 | FK1 |
|---|---|---|---|---|---|---|
| | | 246 μm | 174 μm | 291 μm | 259 μm | 221 μm |
| A’ | Position | | 3717.3 | 3718.7 | | |
| | Width | | 15.4 | 15.9 | | |
| | Intensity ab. | | 0.44 | 0.79 | | |
| | Intensity re. | | 0.02 | 0.04 | | |
| A | Position | 3705.6 | 3703.2 | 3706.3 | 3707.7 | 3709.1 |
| | Width | 16.2 | 16.3 | 15.9 | 15.9 | 16.8 |
| | Intensity ab. | 2.40 | 3.03 | 3.19 | 4.06 | 1.74 |
| | Intensity re. | 0.13 | 0.14 | 0.15 | 0.13 | 0.12 |
| A* | Position | 3691.3 | 3791.2 | 3693.0 | 3694.4 | 3693.6 |
| | Width | 16.3 | 16.3 | 15.9 | 15.9 | 16.8 |
| | Intensity ab. | 5.41 | 5.66 | 4.74 | 6.16 | 3.80 |
| | Intensity re. | 0.29 | 0.27 | 0.23 | 0.20 | 0.26 |
| B | Position | 3674.7 | 3675.4 | 3677.8 | 3680.8 | 3678.5 |
| | Width | 16.3 | 17.5 | 16.8 | 17.2 | 17.2 |
| | Intensity ab. | 2.89 | 3.72 | 2.67 | 5.28 | 2.61 |
| | Intensity re. | 0.15 | 0.18 | 0.13 | 0.17 | 0.18 |
| T | Position | 3671.4 | 3672.4 | 3672.8 | 3671.5 | 3671.2 |
| | Width | 9.3 | 9.2 | 9.3 | 9.3 | 9.3 |
| | Intensity ab. | 0.98 | 1.23 | 2.27 | 2.82 | 0.72 |
| | Intensity re. | 0.05 | 0.06 | 0.11 | 0.09 | 0.05 |
| B* | Position | 3658.5 | 3659.5 | 3661.8 | 3658.8 | 3659.2 |
| | Width | 16.6 | 17.5 | 16.8 | 17.2 | 17.2 |
| | Intensity ab. | 5.06 | 4.93 | 5.25 | 8.97 | 3.93 |
| | Intensity re. | 0.27 | 0.23 | 0.26 | 0.29 | 0.27 |
| G | Position | 3641.3 | 3641.1 | 3640.3 | 3637.8 | 3640.4 |
| | Width | 19.6 | 20.8 | 19.1 | 19.1 | 19.1 |
| | Intensity ab. | 1.39 | 1.57 | 1.28 | 2.54 | 1.26 |
| | Intensity re. | 0.07 | 0.07 | 0.06 | 0.08 | 0.09 |
| H | Position | 3623.0 | 3622.4 | 3624.4 | 3617.0 | 3621.4 |
| | Width | 20.1 | 20.7 | 19.4 | 19.4 | 19.4 |
| | Intensity ab. | 0.48 | 0.52 | 0.39 | 1.02 | 0.45 |
| | Intensity re. | 0.03 | 0.02 | 0.02 | 0.03 | 0.03 |

**Table 7** Final band assignments for the samples of this work. Note that in local environments involving empty *A*-sites (denoted by #) the OH-OH and OH-F configurations cannot be distinguished

| Configuration | | | | Frequency ($cm^{-1}$) | Band |
|---|---|---|---|---|---|
| *M*(1)*M*(1)*M*(3) | *A* | *T*(1)*T*(1) | O(3)-O(3) | | |
| MgMgMg | Na | SiAl | OH-OH | 3710 | A, A’ |
| MgMgMg | Na | SiAl | OH-F | 3692 | A* |
| MgMgAl | Na | SiAl | OH-OH | 3678 | B |
| MgMgAl | Na | SiAl | OH-F | 3660 | B* |
| MgMgMg | □ | SiSi | OH-OH# | 3671 | T |
| MgMgMg | □ | SiAl | OH-OH# | 3642 | G |
| MgMgAl | □ | SiAl | OH-OH# | 3622 | H |

**Table 8** Comparison between the FTIR- and EMPA-derived $^{W}F$ and $^{A}\square$ contents for the studied samples

| Sample | $^{W}F$ | $^{W}F$ | $^{A}\square$ | $^{A}\square$ |
|---|---|---|---|---|
| | FTIR | EMPA | FTIR | EMPA |
| J9698 | 1.36 | 1.41 | 0.07 | 0.08 |
| C2552 | 1.22 | 1.31 | 0.07 | 0.07 |
| Ge408 | 1.28 | 1.14 | 0.10 | 0.21 |
| G405 | 1.24 | 1.46 | 0.10 | 0.09 |
| FK1 | 1.28 | 1.18 | 0.08 | 0.12 |

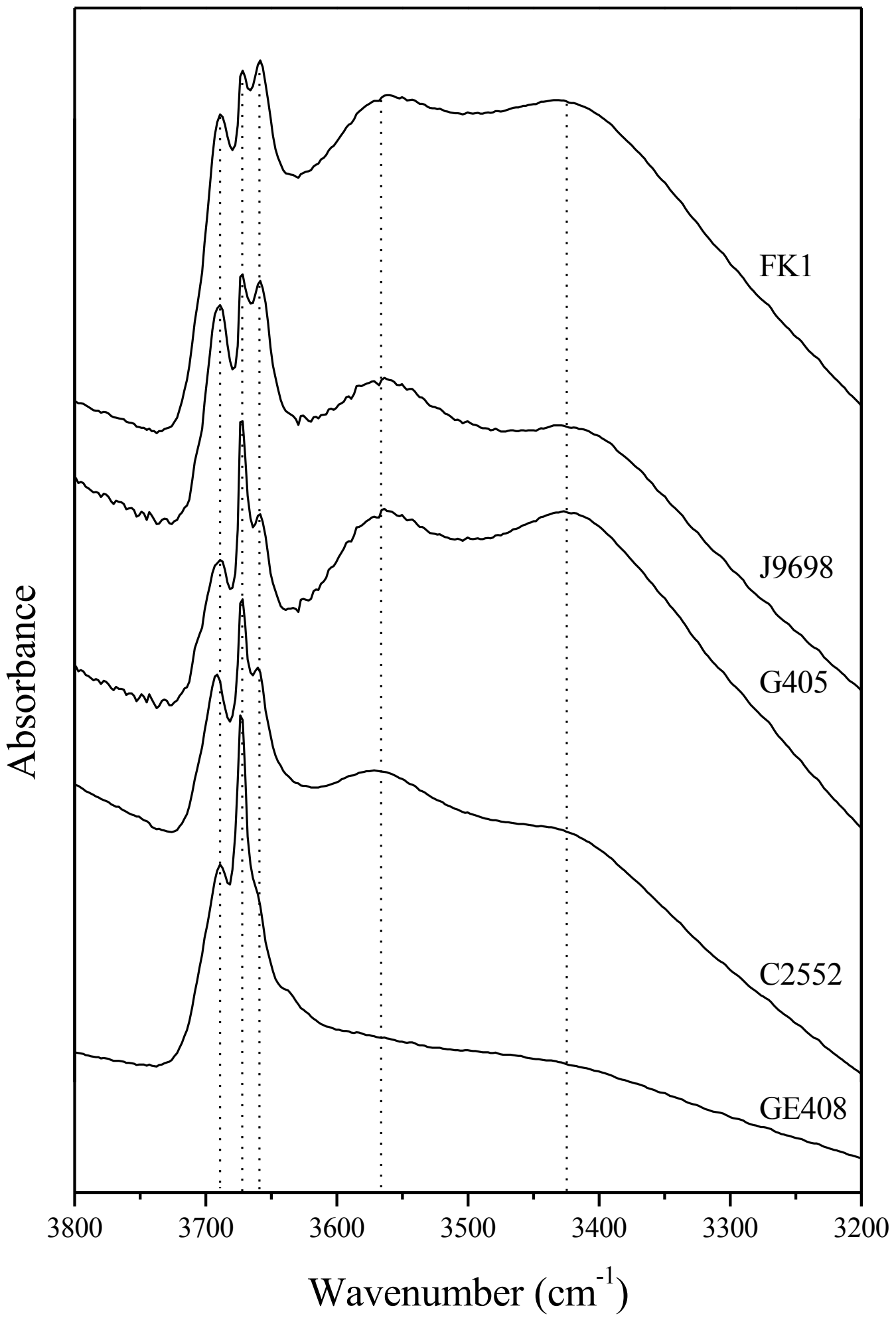

**Fig. 1** IR OH-stretching powder-spectra for the studied fluoro-amphiboles.

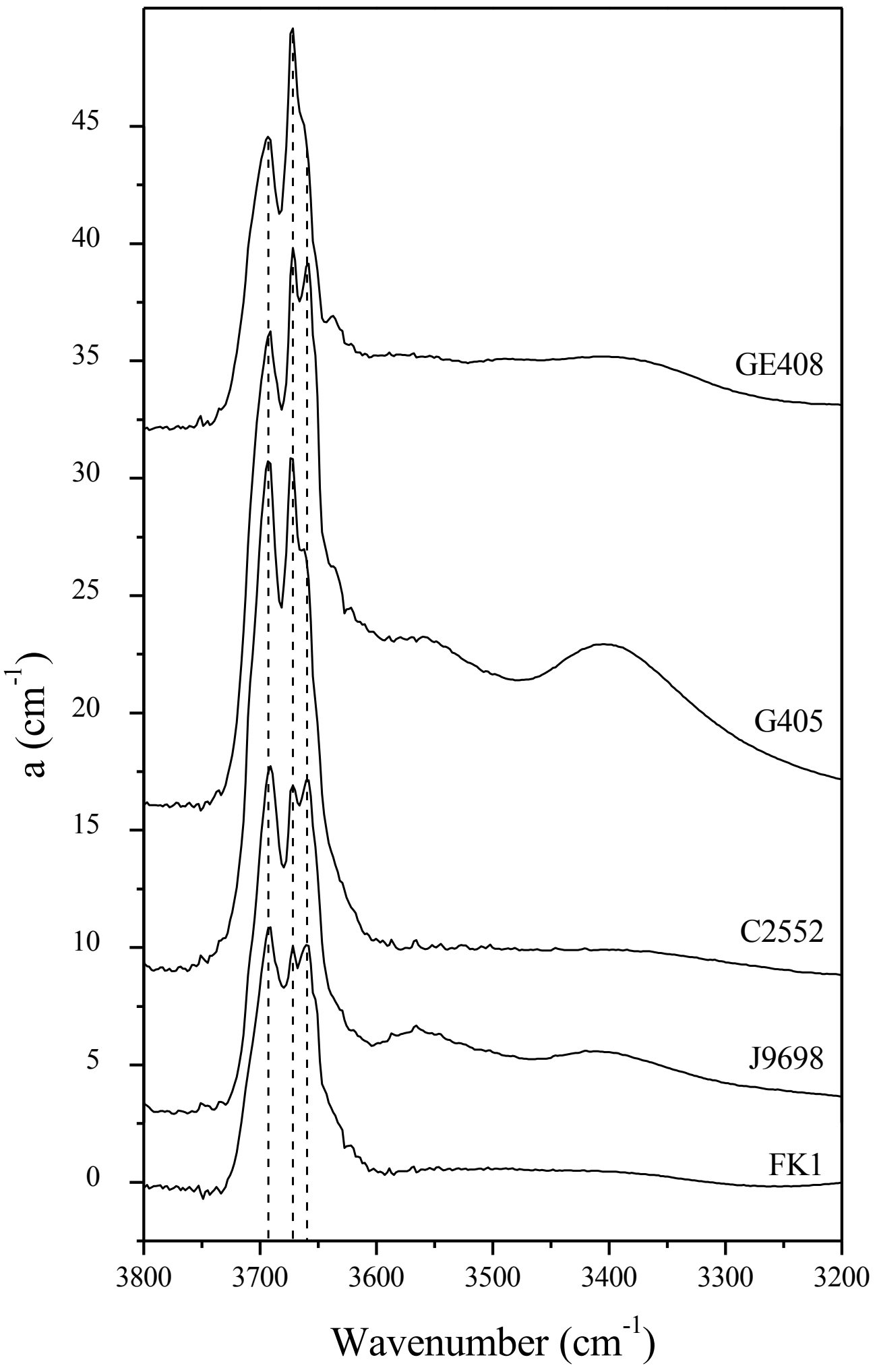

**Fig. 2** Single-crystal IR OH-stretching spectra for the studied fluoro-amphiboles. Absorbance scale normalized to thickness.

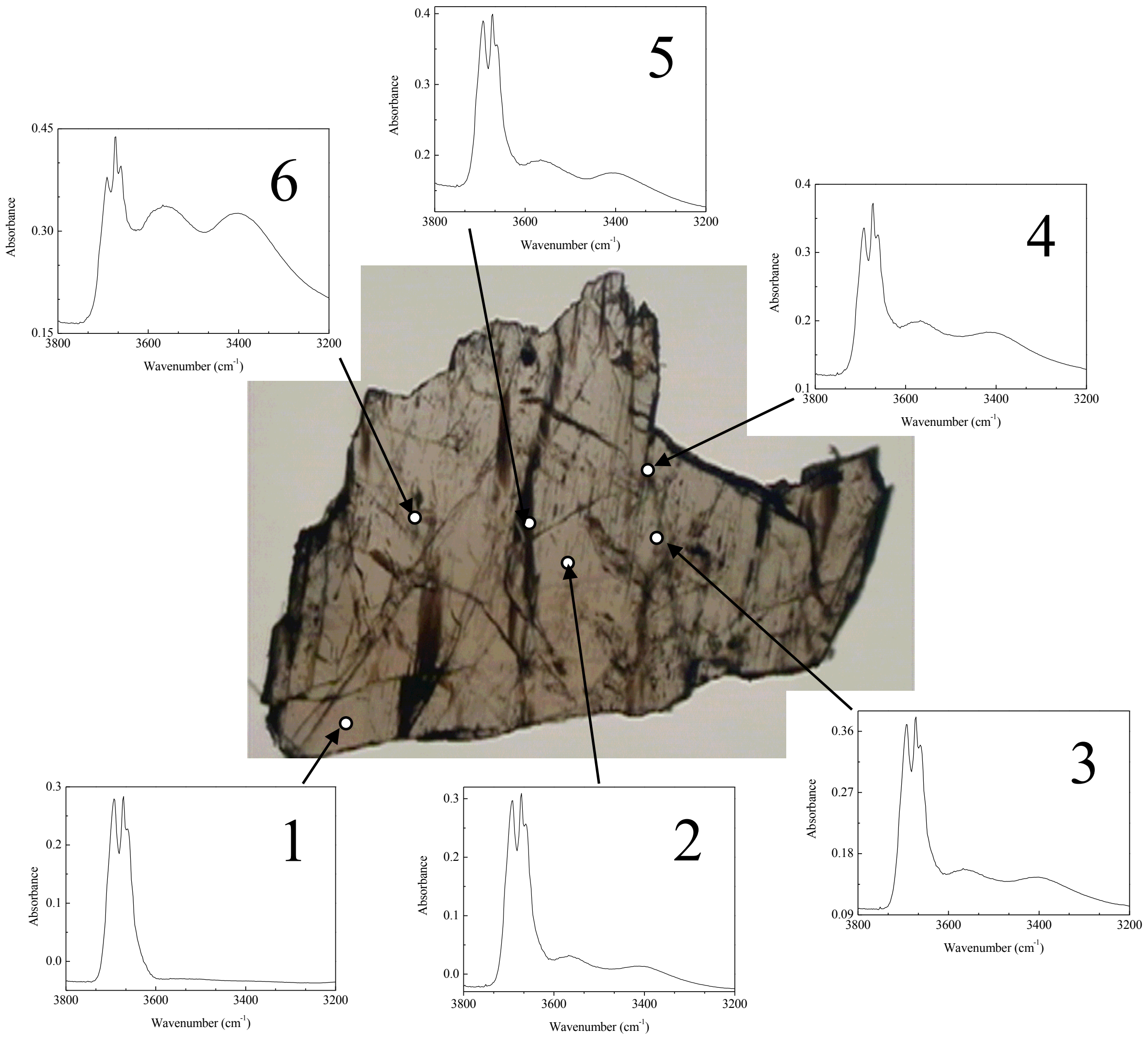

**Fig. 3** Variation of the OH-stretching spectrum as a function of the analysed point. Sample C2552, thickness 174 μm.

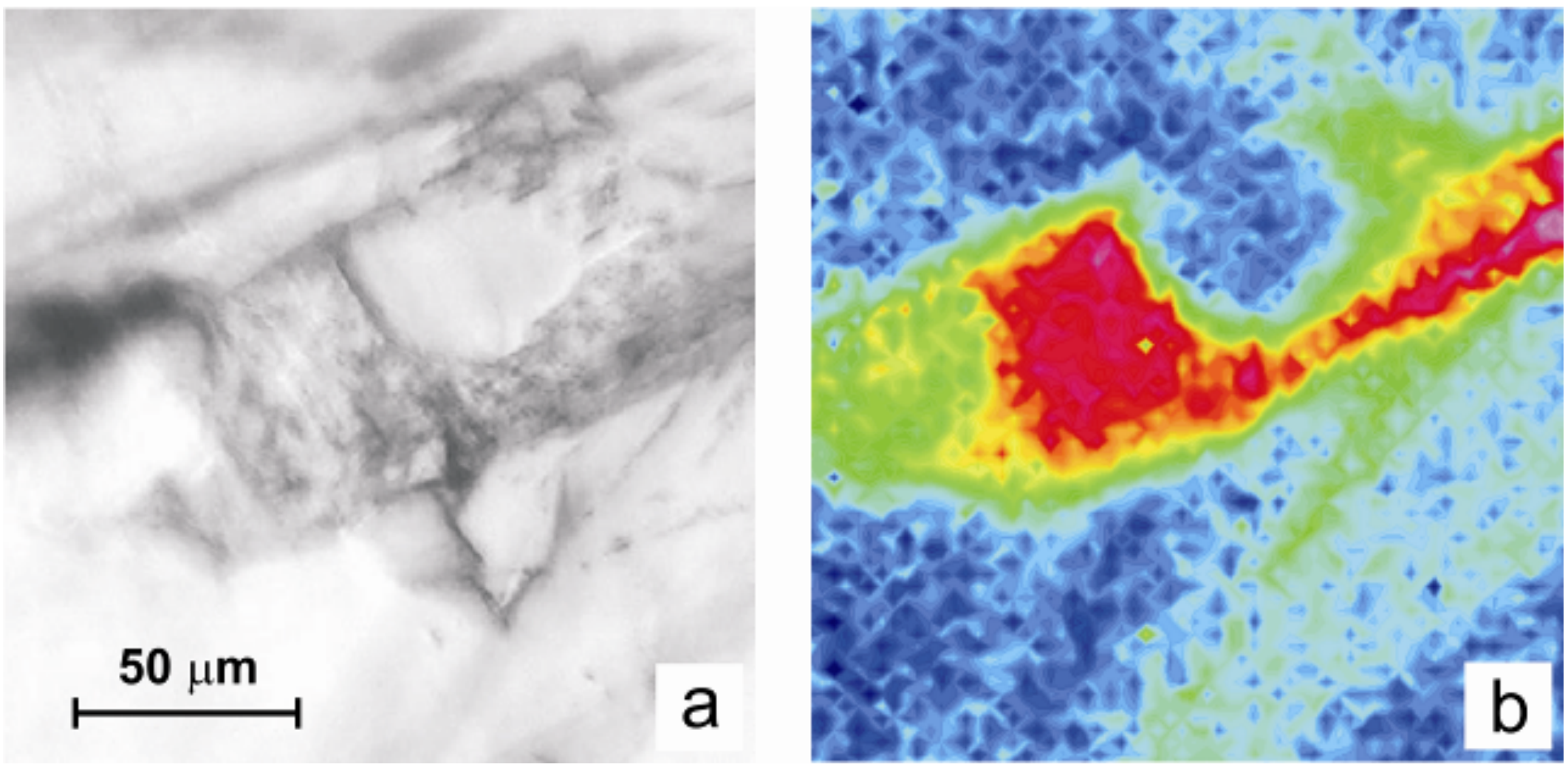

**Fig. 4** Selected FPA image (b) of sample C2552; the corresponding optical image is given in (a). The scale color from blue (minimum) to red (maximum) is proportional to the intensity in the water stretching 3720 - 3200 $cm^{-1}$ region.

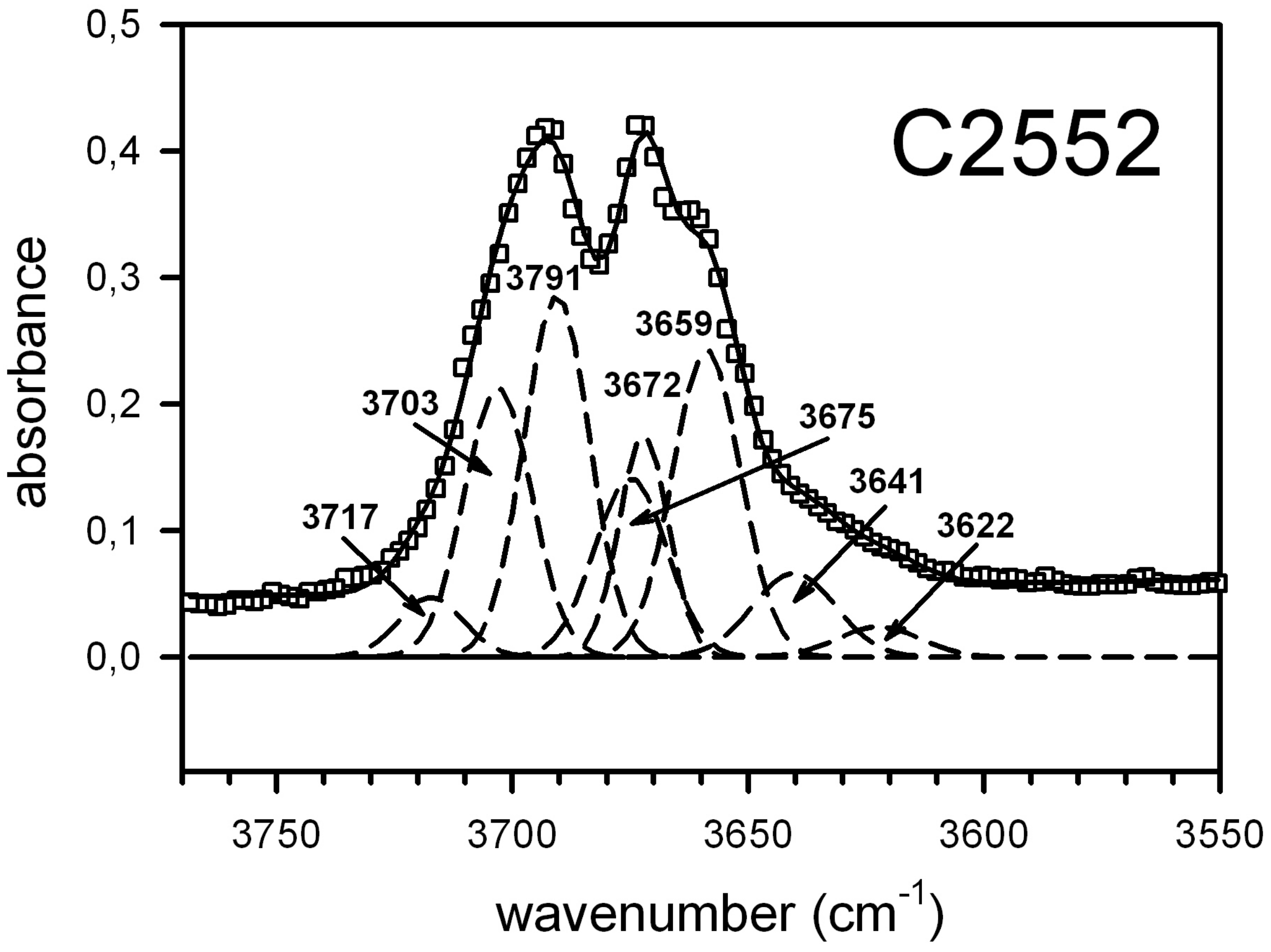

**Fig. 5** The fitted spectrum of sample C2552; open squares: experimental pattern, broken lines: fitted bands, line: resulting envelope.

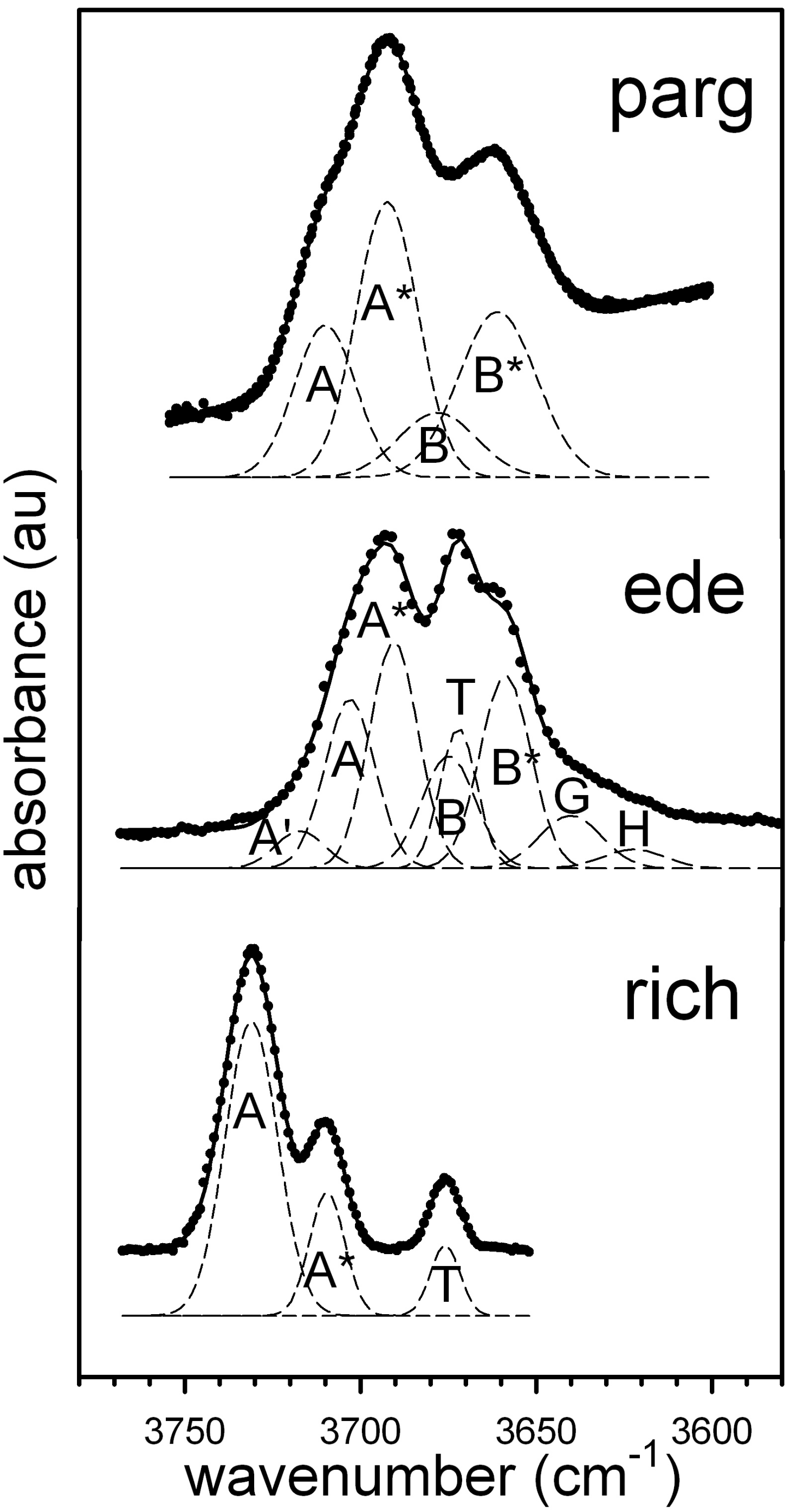

**Fig. 6** The fitted spectrum of Figure 5 (middle) compared with the fitted spectra of synthetic fluoro-pargasite (top, from Robert *et al.,* 2000) and fluoro-richterite (bottom, from Robert *et al.*, 1999), both the synthetic samples samples having F = 1.2 apfu.

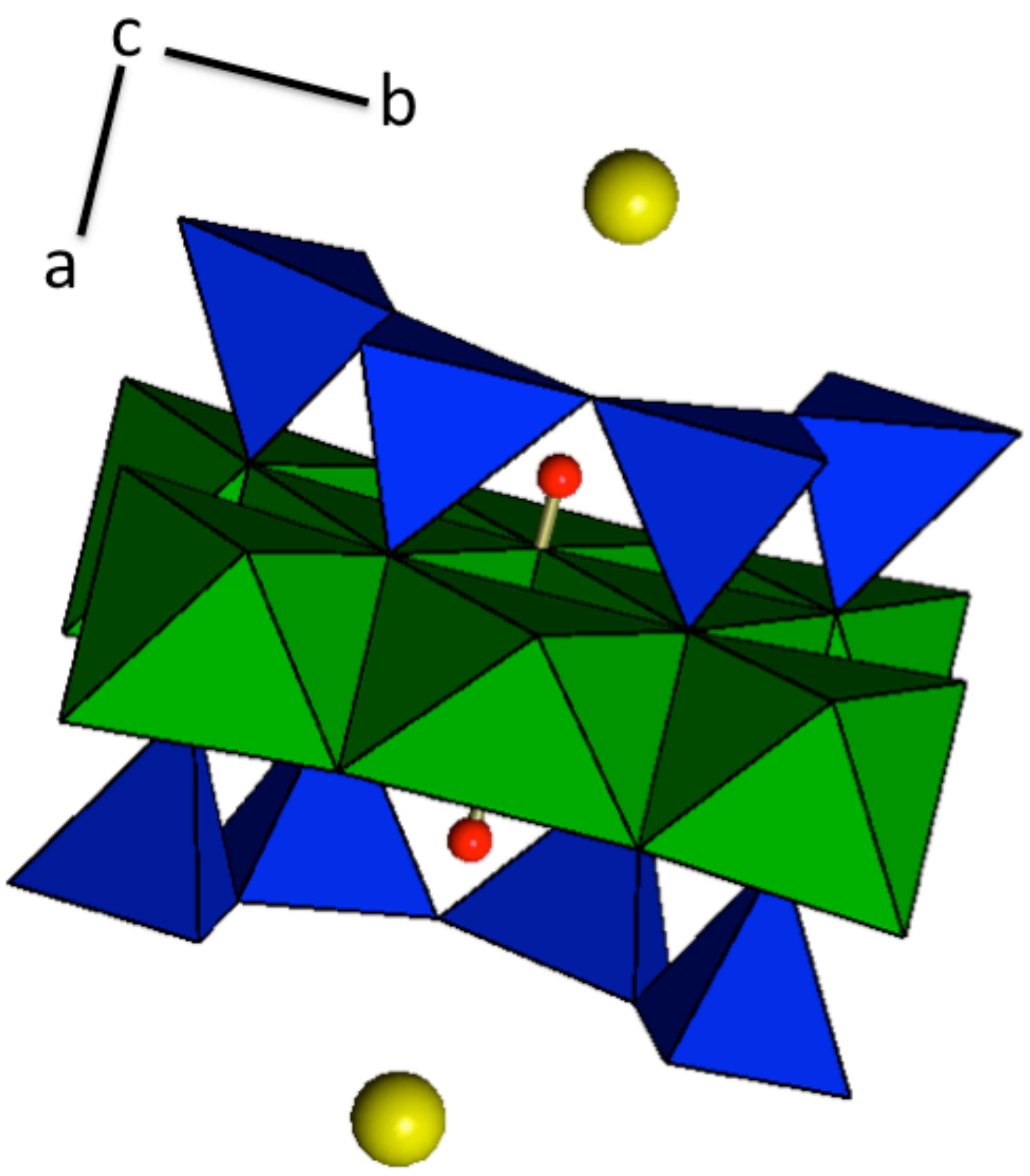

**Fig. 7** A sketch of the *C*2/*m* amphibole structure showing the local environment around the O-H dipole.

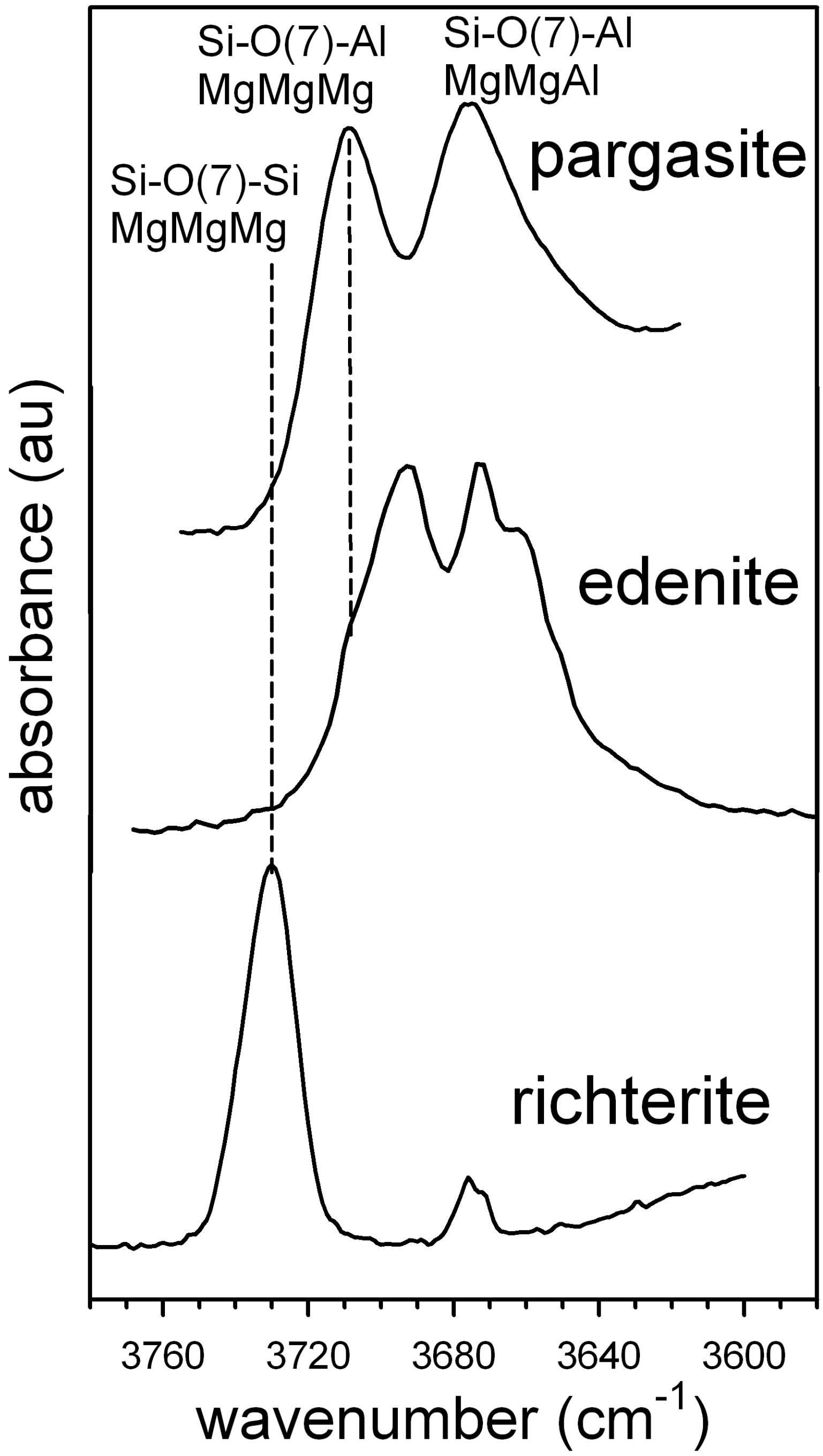

**Fig. 8** The OH-stretching spectra of synthetic pargasite (from Della Ventura *et al*., 1999), fluoro-edenite from Franklin (sample C2552, this work), and synthetic richterite (from Robert *et al*., 1989). The local configurations are schematically given in the figure.

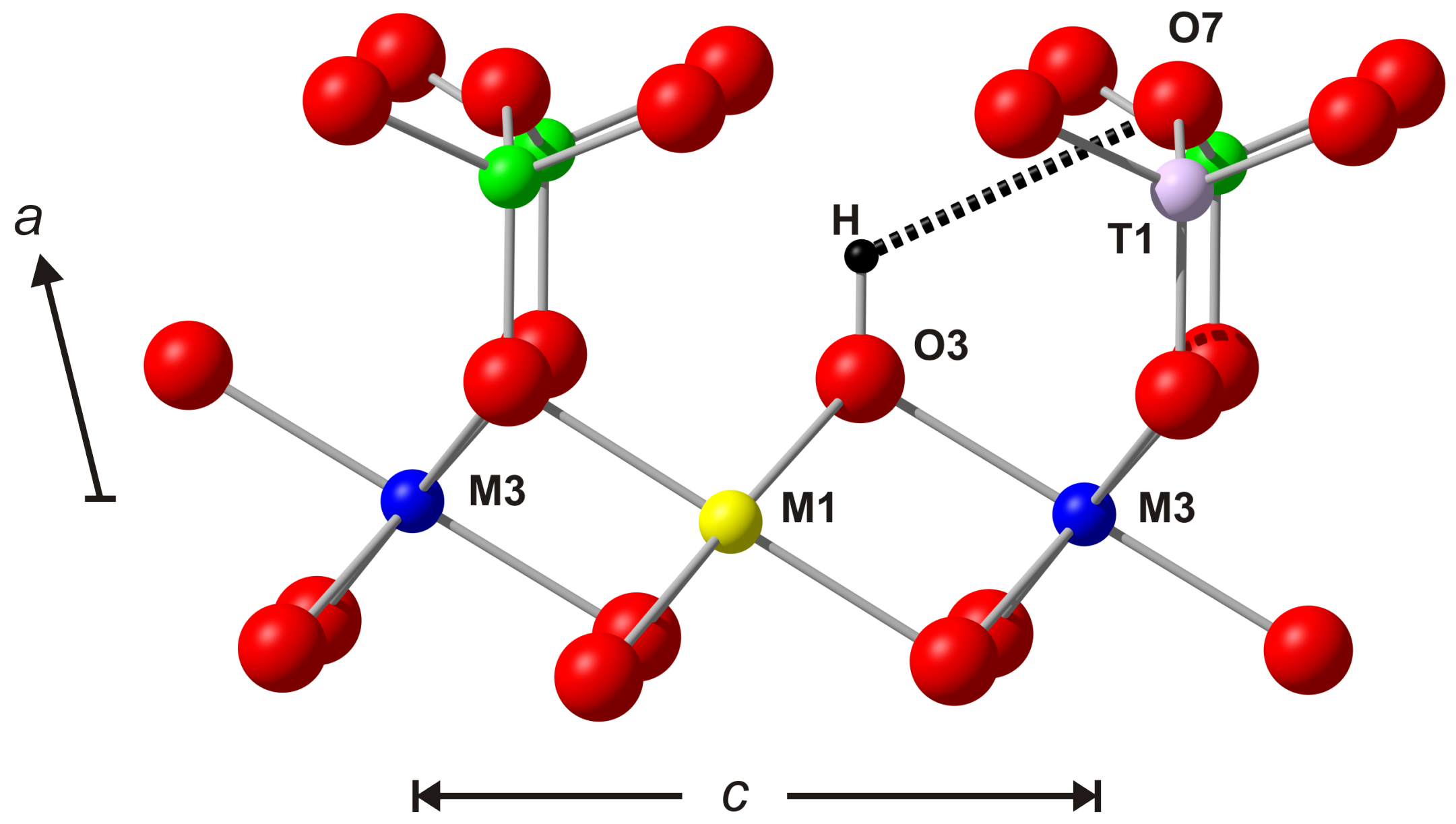

**Fig. 9** The local environment of the H atom in the $C2/m$ amphibole structure view down [010]; the $T(2)$ sites have been omitted for clarity. Green = Si, grey = Al. Modified from Della Ventura *et al*. (1999).

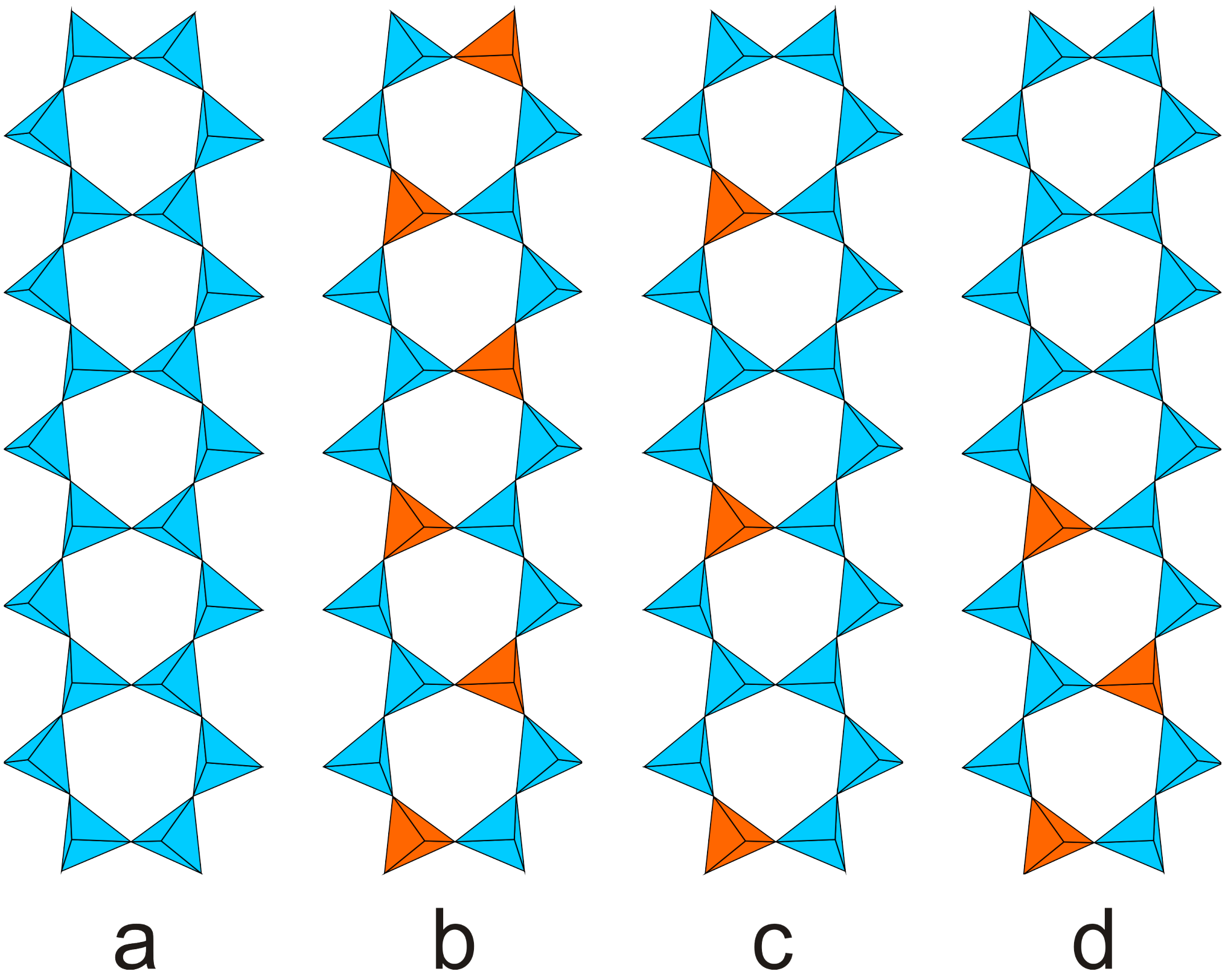

**Fig. 10** Schematic tetrahedral chain arrangements in (a) richterite; (b) pargasite; (c) edenite, regularly alternated; (d) edenite, clustered. For more explanation, see text. Blue tetrahedra = Si, orange tetrahedra = Al. Note that when the *T*(1)-O(7)-*T*(1) linkages have the configuration Al-O(7)-Si there can be orientational disorder relative to the *b* direction.